\documentclass[nofootinbib,preprintnumbers,amsmath,latexsym,array,enumerate,11pt,superscriptaddress]{revtex4}
\usepackage{slashed}
\usepackage{amsmath,amssymb}
\usepackage{amssymb,amsmath}
\usepackage{textcomp}
\usepackage{mathrsfs}
\usepackage{array}
\usepackage{bm,latexsym,amsmath,amssymb,amsfonts,mathrsfs}
\usepackage[dvips]{graphicx}

\usepackage{graphicx}
\usepackage{dcolumn}
\usepackage{bm}
\usepackage{amssymb}
\usepackage{slashed}
\usepackage[utf8]{inputenc}

\usepackage{indentfirst}

\usepackage{xcolor}
\usepackage{dsfont}
\usepackage{ulem}
\usepackage{verbatim}
\usepackage{yfonts}
\usepackage{color}
\usepackage{ifsym}
\usepackage{lipsum}

\usepackage{graphicx}
\usepackage{caption}
\usepackage{appendix}

\usepackage{caption}
\newcommand{\be}{\begin{equation}}
\newcommand{\ee}{\end{equation}}
\newcommand{\beq}{\begin{eqnarray}}
\newcommand{\eeq}{\end{eqnarray}}

\def\g{\gamma}

\def\l{\lambda}

\def\q{\partial}

\def\s{\sigma}
\def\t{\tau}

\def\G{\Gamma}

\def\L{\Lambda}

\def\beq{\begin{eqnarray}}
\def\eeq{\end{eqnarray}}
\newcommand{\gsim}{ \mathop{}_{\textstyle \sim}^{\textstyle >} }

\makeatletter
\newsavebox\myboxA
\newsavebox\myboxB
\newlength\mylenA

\newcommand*\xoverline[2][0.75]{%
    \sbox{\myboxA}{$\m@th#2$}%
    \setbox\myboxB\null
    \ht\myboxB=\ht\myboxA%
    \dp\myboxB=\dp\myboxA%
    \wd\myboxB=#1\wd\myboxA
    \sbox\myboxB{$\m@th\overline{\copy\myboxB}$}
    \setlength\mylenA{\the\wd\myboxA}
    \addtolength\mylenA{-\the\wd\myboxB}%
    \ifdim\wd\myboxB<\wd\myboxA%
       \rlap{\hskip 0.5\mylenA\usebox\myboxB}{\usebox\myboxA}%
    \else
        \hskip -0.5\mylenA\rlap{\usebox\myboxA}{\hskip 0.5\mylenA\usebox\myboxB}%
    \fi}
\makeatother

\begin{document}
\bigskip

\title{Towards gauge unified, supersymmetric hidden strong dynamics}


\author{Cheng-Wei Chiang}
\email[email:]{chengwei@phys.ntu.edu.tw}
\affiliation{Department of Physics, National Taiwan University, Taipei, Taiwan 10617, Republic of China.}
\affiliation{Department of Physics and Center for Mathematics and Theoretical Physics, National Central University, Taoyuan, Taiwan 32001, Republic of China.}
\affiliation{Institute of Physics, Academia Sinica, Taipei, Taiwan 11529, Republic of China.}
\affiliation{Physics Division, National Center for Theoretical Sciences, Hsinchu, Taiwan 30013, Republic of China.}

\author{Sichun Sun}
\email[email:]{sichunssun@gmail.com}
\affiliation{Department of Physics, National Taiwan University, Taipei, Taiwan 10617, Republic of China.}
\affiliation{Institute for Advanced Study, The Hong Kong University of Science and Technology, Clear Water Bay, Kowloon, Hong Kong}

\author{Fang Ye}
\email[email:]{fangye@ntu.edu.tw}
\affiliation{Department of Physics, National Taiwan University, Taipei, Taiwan 10617, Republic of China.}
\affiliation{Department of Physics and Center for Mathematics and Theoretical Physics, National Central University, Taoyuan, Taiwan 32001, Republic of China.}

\begin{abstract}
We consider a class of models with extra complex scalars that are charged under both the Standard Model and a hidden strongly coupled $SU(N)_H$ gauge sector, and discuss the scenarios where the new scalars are identified as the messenger fields that mediate the spontaneously broken supersymmetries from the hidden sector to the visible sector.  The new scalars are embedded into 5-plets and 10-plets of an $SU(5)_V$ gauge group that potentially unifies the Standard Model gauge groups.  The Higgs bosons remain as elementary particles.  In the supersymmetrized version of this class of models, vector-like fermions whose left-handed components are superperpartners of the new scalars are introduced.  Due to the hidden strong force, the new low-energy scalars hadronize before decaying and thus evade the common direct searches of the supersymmetric squarks.  This can be seen as a gauge mediation scenario with the scalar messenger fields forming low-energy bound states.  We also discuss the possibility that among the tower of bound states formed under hidden strong dynamics (at least the TeV scale) one of them is the dark matter candidate, as well as the collider signatures (e.g. diphoton, diboson or dijet) of the models that may show up in the near future.
\end{abstract}



\maketitle

\newpage
\section{Introduction \label{sec:intro}}

Although the Standard Model (SM) in particle physics has achieved great success in phenomenology, there still exist many problems (such as dark matter puzzle, hierarchy issue, and flavor problem) that the SM does not address, leading people to believe the existence of physics beyond the SM (BSM). Popular BSM scenarios include supersymmetry (SUSY), in particular the minimal supersymmetric Standard Model (MSSM) (sometimes with different prefixes, NMSSM, PMSSM, CMSSM~\cite{Cahill-Rowley:2014wba}, etc, by extending the MSSM with additional fields), and composite/little Higgs models with TeV-scale strong dynamics~\cite{ArkaniHamed:2002qy}.

As one of the most fledged scenarios of BSM physics,
SUSY has very good motivations from both theoretical and phenomenological points of view.  In semi-realistic string theories, probably one of the most competent fundamental quantum theories that incorporate gravity, supersymmetries are required for deep reasons.  Supersymmetries not only introduce fermionic degrees of freedom to the otherwise purely bosonic string theories, but also elegantly ensure the stability of the system by eliminating the tachyons (states with negative mass squared that commonly appear in string theories).
Phenomenologists on the other hand particularly favor SUSY at low or intermediate scales.  Such supersymmetric models provide a nice mechanism to explain the large hierarchy between the electroweak (EW) scale and the Planck scale by canceling the quadratic divergence in radiative corrections to the Higgs mass, and therefore save the Nature from being fine-tuned.  Since experimental searches have ruled out a large portion of parameter/model space for low-scale SUSY, people now have to focus on supersymmetric models at or above the TeV scale instead.

In low- or intermediate-scale supersymmetric models, people usually work within the perturbative regime (except for the QCD part) probably just for the cause of simplicity.  In many UV-completed models, however, extra strongly coupled sectors are quite common.  Ref.~\cite{strong-hidden-heterotic:KSTK} shows how such strongly-interacting hidden sectors can arise in heterotic string theories.  For discussions about the strongly coupled sectors in the context of Type-II string theories, one may refer to Ref.~\cite{strong-hidden-typeII:CLS}, a follow-up phenomenological study of the intersecting D-brane models constructed in Refs.~\cite{storng-hidden-typeIIA-intersecting-branes:CSU0107-1,storng-hidden-typeIIA-intersecting-branes:CSU0107-2}.  A pedagogical review about the intersecting D-branes, in particular how the strongly coupled sectors appear, can be found in Ref.~\cite{strong-hidden-D-branes-review:BCLS}. Non-perturbative effects are hard to calculate quantitatively.  The study of the strongly coupled gauge theory and gauge/gravity duality has shed some light on non-perturbative calculations.
The strongly coupled theory can be converted into a weakly coupled sector in the bulk, following the holographic principle.  One example in SUSY breaking is that the visible sector may talk to the strongly coupled hidden sector through messenger fields which are charged under both gauge sectors, as studied in the scheme of holographic gauge mediation~\cite{holographic-gauge-mediation:MSS}.

In this paper, we examine a class of BSM models with an extra strongly coupled hidden $SU(N)_H$ sector, along a similar line of Refs.~\cite{hidden-strong-dynamics:CIY1512,hidden-strong-dynamics:CFHIY1507}. We point out that this type of theories potentially alleviate the little hierarchy problem,
 and can be realized in a gauge-mediated SUSY breaking scenario. New scalars charged simultaneously under both the SM and hidden gauge groups are introduced for the purposes. We explore the possibility of achieving the SM gauge coupling unification at an appropriate scale with the addition of those new particles.  Such models appear much less fine-tuned in the Higgs mass without conflicts with direct search bounds at colliders.  As a bonus, exotic bound states formed under the new strong dynamics appear as various diboson/dijet/diphoton resonances at different scales, the lightest of which may be discovered at colliders soon.

We emphasize that in our models the new (supersymmetric) strong dynamics has entirely different signatures at colliders in comparison with scenarios such as low- or intermediate-scale perturbative SUSY, Goldstone Higgs~\cite{Sun:2013cza} and supersymmetric Goldstone Higgses~\cite{Chang:2006ra}.  The new scalars, despite being around a few hundreds GeV to TeV, can hadronize quickly into exotic mesons and baryons through the new strong dynamics.  In contrast with the standard final state searches for squarks in SUSY, detecting the new scalars requires a different approach since they are confined in the bound states and those new bound states will decay more like pions and Goldstone particles, similar to the resonances in composite Higgs theories.  Different than the composite-Higgs/technicolor scenarios, the Higgs bosons are fundamental particles in our scenarios. Therefore, our models are free from many electroweak precision constraints.

The paper is organized as follows.  In Section~\ref{sec:setup}, we define the gauge groups of the models and provide the particle spectrum in both non-supersymmetric and supersymmetric cases.  The Higgs mass fine-tuning issue is studied in Section~\ref{sec:fine-tuning}.  In Section~\ref{sec:unification}, we explore the conditions on the particle content of our models in order for the SM gauge coupling unification to be achieved at an appropriate scale.  In Section~\ref{sec:bounds-bound-states}, we discuss existing experimental constraints on the new scalar masses, and list possible exotic bound states formed from the new degrees of freedom as well as possible dark matter candidates.  In Section~\ref{sec:750}, we discuss the collider phenomenology of the new mesons arising from our models, discussing in some detail their diphoton/diboson/dijet signatures at the LHC.  The concluding remarks are given in Section~\ref{sec:concl}.

\section{Setup}\label{sec:setup}
\subsection{Particle contents}\label{subsec:non-susy}

We extend the SM by adding complex scalar multiplets charged as fundamentals under a hidden $SU(N)_H$ gauge group with a confinement scale $\Lambda_H\sim\mathcal O(1)$~TeV.  
All the SM fields are neutral under the hidden gauge group.  We choose the SM charges of those new particles as given in Table~\ref{spectrum}, so that one generation of them and their conjugates can be neatly embedded into part of or full irreducible representations of an $SU(5)_V$ gauge group which potentially unifies the SM gauge groups $SU(3)_C\times SU(2)_L\times U(1)_Y$, where the subscript $V$ denotes the visible sector.  We emphasize here that the new particles in the class of models we focus on are all from Table~\ref{spectrum}, but this does not mean for each model new particles have to run the full spectrum Table~\ref{spectrum}. In fact, each set of new particles defines a specific model.  In order for the hidden gauge group $SU(N)_H$ to be confined, the number of particles that are charged under the hidden group cannot be too large. We will check this issue in a specific model (\ref{non-susy-example}) later.

Explicitly, we have for the messenger fields:
\begin{eqnarray}
\bar{\textbf{5}}=\left(\begin{array}{c}\label{rep_5bar}
\widetilde{D}^{\dagger}\\
\widetilde{L}
\end{array}\right),
\end{eqnarray}
\begin{eqnarray}
\textbf{10}=\left(\begin{array}{ccccc}\label{rep_10}
0 & \widetilde{U'}_{3}^{\dagger} & \widetilde{U'}_{2}^{\dagger} & \widetilde{Q}_{U1} & \widetilde{Q}_{D1}\\
-\widetilde{U'}_{3}^{\dagger} & 0 & \widetilde{U'}_{1}^{\dagger} & \widetilde{Q}_{U2} & \widetilde{Q}_{D2}\\
-\widetilde{U'}_{2}^{\dagger} & -\widetilde{U'}_{1}^{\dagger} & 0& \widetilde{Q}_{U3} & \widetilde{Q}_{D3}\\
-\widetilde{Q}_{U1} & -\widetilde{Q}_{U2} & -\widetilde{Q}_{U3} & 0 & \widetilde{E}^{\dagger}\\
-\widetilde{Q}_{D1} & -\widetilde{Q}_{D2} & -\widetilde{Q}_{D3} & -\widetilde{E}^{\dagger} & 0
\end{array}\right),
\end{eqnarray}
where
\begin{eqnarray}
\widetilde{D}^{\dagger}=\left(\begin{array}{c}
\widetilde{D}^{\dagger}_{1}\\
\widetilde{D}^{\dagger}_{2}\\
\widetilde{D}^{\dagger}_{3}\end{array}\right),\quad \widetilde{L}=\left(\begin{array}{c}
\widetilde{L}_N\\ 
\widetilde{L}_E\end{array}\right).
\end{eqnarray}

The wide tilde indicates those new fields are scalars~\footnote{We put a prime in $\widetilde{U'}^{\dagger}$ here, to distinguish it from another multiplet to be introduced in the supersymmetrized setup and to be embedded in another $SU(5)_V$ $10$-plet.}. 
For convenience, we will call the particles associated with the messenger fields as hidden scalars. At this stage, we have not fixed the number of generations for each multiplet. In other words, in a specific model with a low energy effective theory, a multiplet may be present or absent, and the new multiplets may come in complete or incomplete $SU(5)_V$ 5/10-plets. As will be shown in Section~\ref{sec:unification}, from the gauge coupling unification point of view, actually incomplete GUT representations are required.
\begin{table}
\begin{tabular}{cccccc}
\hline\hline
 & $SU(N)_H$ & $SU(3)_C$ & $SU(2)_L$ & $U(1)_Y$&$U(1)_{EM}$\\
 \hline 
 \vspace{-0.5cm}
 \\
 $\widetilde{Q}=(\widetilde{Q}_U,\,\widetilde{Q}_D)^T$ &$\textbf{N}$&  $\textbf{3}$ & $\textbf{2}$ &$1/6$& $2/3,\,-1/3$\\
 $\widetilde{U'}^{\dagger}$ & $\textbf{N}$ & $\overline{\textbf{3}}$ & $\textbf{1}$ & $-2/3$&$-2/3$\\
 $\widetilde{D}^{\dagger}$ &$\textbf{N}$&  $\bar{\textbf{3}}$ & $\textbf{1}$ & $1/3$&$1/3$\\
 $\widetilde{L}=(\widetilde{L}_N,\,\widetilde{L}_E)^T$ & $\textbf{N}$& $\textbf{1}$ & $\textbf{2}$ & $-1/2$&$0,\,-1$\\
 $\widetilde{E}^{\dagger}$ & $\textbf{N}$& $\textbf{1}$ & $\textbf{1}$ & $1$&$1$\\
\hline\hline
\end{tabular}
\caption{Representations of some new messenger fields.  The dagger denotes the Hermitian conjugate. The electric charge is related to the hypercharge through $Q_{EM}=T_3+Y$.}\label{spectrum}
\end{table}

In addition to the new fields in Table~\ref{spectrum}, we consider a Higgs sector with two Higgs doublet fields, both neutral under $SU(N)_H$, as shown in Table~\ref{Higgs}. Similar to the Two-Higgs Doublet Model (2HDM), one linear combination of the electrically neutral Higgs bosons is identified as the $125$-GeV Higgs boson.  There are a few distinct scenarios, depending on how the two Higgs fields couple with SM fermions.  We emphasize here that in our setup the Higgs fields are fundamental, different from the usual technicolor or composite/little Higgs models in which the Higgses are (pseudo)-Goldstone bosons.  Therefore, our models are exempt from the usual constraints of those models.
\begin{table}
\centering
\begin{tabular}{cccccc}
\hline\hline
 & $SU(N)_H$ & $SU(3)_C$ & $SU(2)_L$ & $U(1)_Y$&$U(1)_{EM}$\\
 \hline 
 $H_u=( H^+_u,\, H^0_u)^T$ & $\textbf{1}$ & $\textbf{1}$ & $\textbf{2}$ &$1/2$&$1,\,0$\\
 $ H_d=( H^0_d,\,H^-_d)^T$ &$\textbf{1}$  & $\textbf{1}$ & $\textbf{2}$ & $-1/2$&$0,\,-1$\\
 \hline\hline
\end{tabular}
\caption{Representations of the Higgs doublets.}\label{Higgs}
\end{table}

\subsection{Supersymmetrized Version}

In this subsection, we briefly comment on the supersymmetric version of the above-mentioned setup.  We will assume that the hidden gauge fields are in the confined phase (This can be achieved for $2\leq N$ as we will show later).  Since the hidden gauge field take no SM charges, we will focus on the messenger sector.  The supersymmetric version of the setup discussed previously includes the MSSM in the visible sector, chiral supermultiplets charged as in Table~\ref{spectrum}  (and their conjugates) and vector supermultiplets associated with the gauge fields in the hidden sector.  We assume that all the MSSM gauginos and Higgsinos are either sufficiently heavy or completely decoupled from the SM so that bounds from the neutralino dark matter direct detection or the direct collider searches ~\cite{Low:2014cba,Bramante:2015una} can be evaded.
In order not to introduce anomalies, we include vector-like fermions whose left-handed components are superpartners of the scalars in Table~\ref{spectrum} and whose right-handed components take conjugate charges of those in Table~\ref{spectrum}.  We denote the left-handed fermions as $Q$, $U'^{\dagger}$, $D^{\dagger}$, $L$, and $E^{\dagger}$ respectively, and their superpartners are those with widetildes as listed in Table~\ref{spectrum}. The  corresponding right-handed fermions and their scalar superpartners are denoted by fields with bars, e.g. $\bar{Q}$ and $\widetilde{\bar{Q}}$.~
\footnote{Note that in Subsection~\ref{subsec:non-susy} we have not fixed the number of each type of field in Table~\ref{spectrum}.  In the supersymmetric version of our setup, each vector-like hidden fermion is associated with two complex scalars in the same representation.}.

In order to allow a Yukawa coupling between the messenger fields and one Higgs doublet field in the superpotential, we introduce another left-handed chiral multiplet, embedded in another $SU(5)_V$ 10-plet and whose scalar component are charged as in Table~\ref{spectrum-2}. 
$\widetilde{U}^{\dagger}$ is embedded in a 10-plet which takes the same visible charges as $\widetilde{U'}^{\dagger}$, $\widetilde{Q}$ and $\widetilde{E}^{\dagger}$ but is antifundamental under the hidden gauge group $SU(N)_H$, comparing to $\widetilde{U}^{\dagger}$. 
\begin{table}[h]
\centering
\begin{tabular}{cccccc}
\multicolumn{6}{c}{}\\
\hline\hline
 & $SU(N)_H$ & $SU(3)_C$ & $SU(2)_L$ & $U(1)_Y$&$U(1)_{EM}$\\
 \hline 
 \vspace{-0.5cm}
 \\
 $\widetilde{U}^{\dagger}$ & $\overline{\textbf{N}}$ & $\overline{\textbf{3}}$ & $\textbf{1}$ & $-2/3$&$-2/3$\\
 \hline\hline
\end{tabular}
\caption{Representations of an additional new messenger field.  The dagger denotes the Hermitian conjugate. The electric charge is related to the hypercharge through $Q_{EM}=T_3+Y$. }\label{spectrum-2}
\end{table}

The Yukawa-type superpotential takes the form:
\begin{eqnarray}\label{superpotential}
W_\text{mess2}\supset Y_U  & \Phi_{\widetilde{U}^{\dagger}}\Phi_{\widetilde{Q}} \Phi_{H_u} ~,
\end{eqnarray}
where $\Phi_{\widetilde{Q}}$, $\Phi_{\widetilde{U}^{\dagger}}$ and $\Phi _{H_u}$ refer to the left-handed (holomorphic) supermultiplets of the corresponding hidden scalars $\widetilde{Q}$, $\widetilde{U}^{\dagger}$ and the Higgs $H_u$, respectively. The coupling of the right-handed superfields can be written down in a similar manner. The hidden scalars $\widetilde{U}$ and $\widetilde{Q}$ thus obtain interactions with $H_u$ in the scalar potential
\begin{eqnarray}\label{messscalar-mass1}
V\supset -|Y_U|^2\left(H_u^{\dagger}H_u\widetilde{Q}^{\dagger}\widetilde{Q}+H_u^{\dagger}H_u\widetilde{U}^{\dagger}\widetilde{U}+\widetilde{Q}^{\dagger}\widetilde{Q}\,\widetilde{U}^{\dagger}\widetilde{U}\right) ~,
\end{eqnarray}
while the hidden fermions $Q$ and $U$ acquire a Yukawa coupling to $H_u$ in the Lagrangian
\begin{eqnarray}\label{messfermion-mass1}
\mathcal{L}_{\rm Yukawa}\supset -Y_U U^{\dagger}H_u Q + {\rm c.c.} ~.
\end{eqnarray}

Although our scenario does not depend much on the way of SUSY-breaking mediation, it is naturally a gauge mediation scenario with messenger fields. If we assume that SUSY in the hidden sector is spontaneously broken by a nonvanishing F-term vacuum expectation value (VEV) $\langle F \rangle$, causing mass splitting between the fermion and scalar in any hidden chiral supermultiplet, the SUSY breaking effects are then transmitted to the visible sector via loops of the messenger particles and modify the MSSM gaugino masses, while the masses of SM gauge bosons remain intact due to the gauge symmetries.  The usual messenger superpotential in which a messenger multiplet $\Phi_\phi$ couples to a singlet multiplet $\Phi_S$ is,
\begin{eqnarray}\label{superpotential1}
W_{mess1}=Y_S\Phi_S\Phi_\phi \bar{\Phi},
\end{eqnarray}
with the F-term corresponding to $\Phi_S$, $<F_S>\neq 0$. In addition to many gauge mediation model with above $W_{mess1}$, there is another messenger superpotential (\ref{superpotential}). Therefore, messenger fermions and scalars receive contributions from both (\ref{superpotential}) and (\ref{superpotential1}). In particular, the scalar potential includes 
\begin{eqnarray}\label{V}
V\ni \biggl|\frac{\partial (W_{mess1}+W_{mess2})}{\partial \phi}\biggr|^2+\biggl | \frac{\partial (W_{mess1}+W_{break})}{\partial S}\biggr|^2,
\end{eqnarray}
where both SUSY and SUSY-breaking contribution to the messenger scalar masses are considered.  Generically the messenger scalars in Table~\ref{spectrum} are not in their mass eigenstates. For mass eigenstates with very large eigenvalues, they decouple from the low energy spectrum and may be identified as ``those missing in the complete gauge unified multiplets", as we will see in Chapter~\ref{sec:unification}. Furthermore, (\ref{V}) may not be the full scalar potential due to the existence of the hidden strong dynamics. The additional messenger superpotential (\ref{superpotential}) and hidden strong force make it possible that the messenger scalar masses are smaller than their fermionic superpartners, which is reverse to the case in MSSM. Explicit derivation of the messenger mass eigenvalues requires the knowledge of the complete messenger spectrum (which is to be determined in Chapter~\ref{sec:unification}) and more details about the hidden sector. We leave it to the future work. In the following we assume that the messenger scalars are much lighter than the messenger fermions and use  $M_{\text{mess}}$ to denote a messenger scalar mass.
Through SUSY-breaking mediation to the visible sector, the SM squarks obtain soft SUSY-breaking masses:
\be
m_\text{soft} \sim \frac{\alpha}{4\pi} \frac{\langle F \rangle}{M_\text{mess}} ~,
\ee
where $\alpha/4\pi$ is the loop factor for gauge mediation. The lightest messenger mass $M_\text{mess}$ in our set-up is relatively lower than that in the standard gauge mediation scenarios.  Here $M_\text{mess}$ can be a few hundred GeV. The vector-like hidden fermions are allowed to have SUSY-breaking mass terms
\be
M_q (Q \bar{Q}+U^{\dagger}\bar U^{\dagger}+...) +c.c.~,
\ee
where we have denoted the messenger fermions schematically as $M_q$.  Both $M_q$ and $m_\text{soft}$ are around the SUSY breaking scale at a few TeV at least, which implies $\sqrt{\langle F\rangle } \gsim 10^4$ GeV.

\section{Fine Tuning of the Higgs Mass}\label{sec:fine-tuning}

The fine tuning in the Higgs potential in general comes from the quadratic dependence in quantum corrections involving two disparate energy scales ({\it i.e.}, the weak scale and the grand unified or Planck scale).  In our setup, the Higgs sector consists of two Higgs doublets, as shown in Table~\ref{Higgs}.  
The SM Higgs boson $H$  with a mass of 125 GeV is a linear combination of the neutral components of these two doublets.  A supersymmetric spectrum ensures the quadratic divergence cancels within the superpartners, leaving SUSY-breaking logarithmic pieces.

The top Yukawa coupling $y_t$ in the SM fermion sector yields the dominant contributions to $m^2_{H_u}$ at one-loop level.  The loop corrections to $m^2_{H_u}$ can thus be used as a parameter to define ``naturalness''~\cite{Kitano:2006gv,Perelstein:2007nx}.  In our models, we can calculate the contribution from the SM sector and the strongly coupled hidden sector as follows if we assume a high-scale supersymmetric spectrum.  The contribution from the SM chiral multiplets reads:
\begin{align}
\delta m^2_{H_u} & \supset \frac{3g_2^2 }{8 \pi^2}\left(M_t^2\,\text{Ln}\frac{\Lambda^2}{M_{t}^2}-M_{\widetilde{t}}^2\,\text{Ln}\frac{\Lambda^2}{M_{\widetilde{t}}^2}\right)
\end{align}
where $M_t$ and $M_{\widetilde{t}}$ denote the top and stop masses, respectively, and we assume the SUSY breaking scale is much heavier than $M_t,\, M_{\widetilde{t}}$, and around 30 TeV.
Note that here we have not taken into consideration of different SUSY-breaking schemes and RG running from the messenger scale to the stop scale.  We simply take the dominant contribution from the mass splitting of the top quark and the top squark. We also do not distinguish the flavor eigenstate and mass eigenstate, and simply assume all mass here coming from the mass eigenstate.
\begin{figure}[t]
\includegraphics[width=10cm]{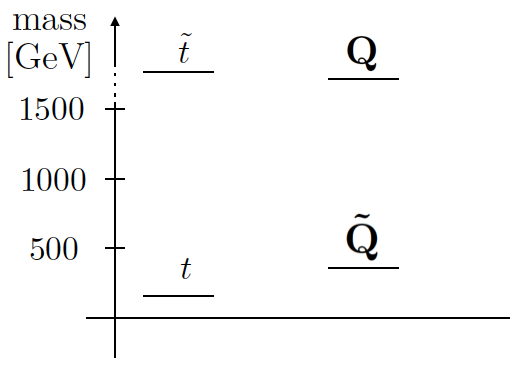}
\caption{A schematic mass spectrum of the top quark, the hidden scalars and their superpartners.  The lightest hidden scalar mass is assumed to be around 300 GeV or above to evade the electroweak precision and Higgs data bounds, while their lightest mesonic bound states to be around the TeV scale. The MSSM stop $\widetilde{t}$ and fermions from the hidden sector multiplets $Q$ are at least around a few TeV. }\label{fig:spectrum}
\end{figure}
\begin{figure}[h!]
\includegraphics[width=8cm]{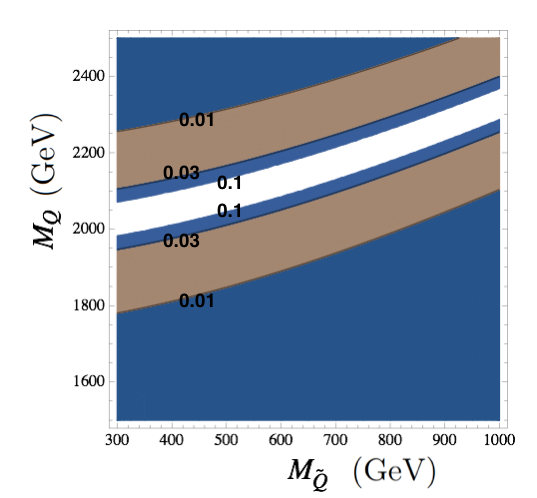}
\includegraphics[width=8cm]{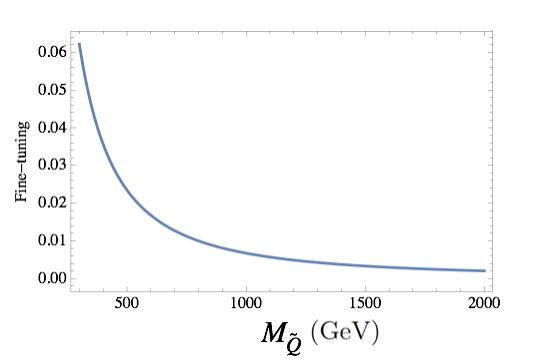}
\caption{Left panel: contours of fine-tuning parameter $\Delta$ in the hidden scalar mass and associated hidden fermion mass plane, with the mass of stop from MSSM being fixed at 3 TeV, and $\Lambda \sim 50$ TeV. $\Delta$ diverges as logarithmic part of one-loop corrections to the Higgs mass completely cancel in the white region.  Right panel: $\Delta$ as a function of the hidden scalar mass in the very-high scale susy case, in which the MSSM stop $\widetilde{t}$ and hidden fermion $Q$ contributions are not considered. We set $N=2$ in both plots, as suggested in Section~\ref{sec:unification}.}\label{fig:fine-tuning}
\end{figure}

Considering the contribution from the hidden sector, dominantly from the mass splitting between the hidden scalar and its associated fermion~\footnote{Note that we have specified neither the number of generations of each multiplet given in Table~\ref{spectrum} nor the name of each ``flavor.''  For the superpotential in Eq.~(\ref{superpotential}), the hidden fermion refers to a $Q$ or $U$. The $\widetilde{Q}$ and $\widetilde{U}^{\dagger}$ components corresponding to the left-handed superpartners play a similar role as the SM stops in the fine-tuning.}, we have
\begin{align}
\delta m^2_{H_u} & \supset\frac{3Y_U^2 N_c }{4 \pi^2}\left(M_{Q}^2\, \text{Ln}\frac{\Lambda^2+M_{Q}^2}{M_{Q}^2}-M_{\widetilde{Q}}^2\, \text{Ln}\frac{\Lambda^2+M_{\widetilde{Q}}^2}{M_{\widetilde{Q}}^2}\right )
\end{align}
where the extra factor of $2$ comes from the vector-like generations, $N$ is the hidden multiplicity due to the $SU(N)_H$ gauge group, and we have ignored a spectrum-dependent overall factor from transforming the messenger fields to their mass eigenstates. $M_Q$ and $M_{\widetilde{Q}}$ refer to the mass eigenvalues of the messenger fermion and of the messenger scalar respectively.

With $M_{Q} \sim M_{\widetilde{t}}$ at a few TeV scale within the reach of next-generation colliders and $M_{\widetilde{Q}}$ around the TeV scale in charge of mainly canceling the quadratic divergence from the SM top quark, we define a naturalness parameter
\beq
\Delta^{-1} =  \left| \frac{2\delta m^2_{H_u}}{m_H^2} \right| ~.
\eeq
We plot a schematic mass spectrum of the top quark, the hidden scalar and their superpartners in Figure~\ref{fig:spectrum}.  The contour of $\Delta$ on the $M_{\widetilde{Q}}$-$M_{Q}$ plane is shown in the left panel of Figure~\ref{fig:fine-tuning}.  In the right panel, we assume a very high-scale susy spectrum by setting the stop and hidden fermion masses to infinity.  Here we have only considered quadratic divergence coming from the top related sector at the one-loop level and set $N$ to 2 in both plots, as suggested in Section \ref{sec:unification}. The quadratic contributions to the Higgs mass coming from the other SM particles are much smaller and presumably can be canceled by introducing corresponding new particles at higher energies.  It is noted that with higher multiplicities, the fine-tuning problem becomes worse.  Besides, a perfect cancellation requires an {\it ad hoc} relation between $y_t$ and $Y_U$.

\section{Gauge Coupling Unification}\label{sec:unification}

It is tempting to see whether the new particles together with the SM particles (and possibly with some other particles) can be embedded into a larger gauge group $SU(5)_{V}\times SU(N)_H$.  People have studied extensively the Grand Unified Theories (GUT) and their applications in phenomenology~\cite{GUT-pheno:DRW82,GUTzilla-DM:HLL2016}. In this section we explore under what conditions the SM gauge couplings can be unified at a GUT scale $M_{\rm GUT}$. For simplicity, we assume that there is no intermediate stage of the symmetry breaking $SU(5)_V\to SU(3)_C\times SU(2)_L\times U(1)_Y$, and that the visible group is spontaneously broken down to the SM gauge group due to ``GUT-Higgs" scalars in the adjoint representation $\textbf{24}$ acquiring a nonvanishing VEV as usual:
\begin{eqnarray}
\langle \textbf{24} \rangle=\text{diag}\left(2,\,2,\,2,\,-3,\,-3\right) v ~.
\end{eqnarray}

At one-loop level, we want the SM gauge couplings $\alpha_a = g_a^2 /(4\pi)$~\footnote{$g_2$ and $g_3$ correspond to $SU(2)_L$ and $SU(3)_C$ gauge coupling constants, respectively. $g_1=\sqrt{\frac{5}{3}}g_Y$, where $g_Y$ is the hypercharge coupling constant.} to be unified at an appropriate GUT scale,
\begin{eqnarray}\label{gauge-coupling-unification}
\alpha_3(M_{\rm GUT})=\alpha_2(M_{\rm GUT})=\alpha_1(M_{\rm GUT})\equiv\alpha_{\rm GUT} ~,
\end{eqnarray}
with the following additional conditions:

\begin{enumerate}
\item The couplings remain within the perturbative regime at the GUT scale, {\it i.e.},
\begin{eqnarray}
\label{condi:perturbative} 
0 < \alpha_{\rm GUT} < 1.
\end{eqnarray}
\item The GUT scale stays within an appropriate range.  We require that the GUT scale is lower than the fundamental string scale $M_s$ (which is lower than the reduced Planck mass $M_P$). On the other hand, the GUT scale should be high enough not to incur a fast proton decay. Since the quarks and leptons are in the same GUT representation, proton can decay via higher dimensional baryon number violating operators. Dimensional analysis indicates the proton lifetime as $\tau _p\sim M^4_{X}/m_p^5$, where $m_p$ is the proton mass and $M_X\sim M_{\rm GUT}$ is the mass of GUT gauge bosons that acquire mass when the GUT group is broken. A detailed calculation~\cite{split-susy} shows that the experimental Super-Kamiokande limit~\cite{SKlimit} $\tau_p>5.3\times 10^{33}\,\text{yr}$ requires
\begin{eqnarray}\label{condi:proton-decay}
M_{\rm GUT} > 
6\times 10^{15}\,\text{GeV} \times
\left(\frac{\alpha_{\rm GUT}}{1/35}\right)^{1/2}\left(\frac{\alpha _N}{0.015\,\text{GeV}}\right)^{1/2}\left(\frac{A_L}{5}\right)^{1/2} ~,
\end{eqnarray}
where the operator renormalization factor $A_L$ and the hadronic matrix element $\alpha_N$ are respectively $5$  and $0.015$~GeV from a lattice calculation~\cite{lattice}.
\end{enumerate}

The unification condition (\ref{gauge-coupling-unification}) leads to
\begin{eqnarray}
\alpha_{GUT}^{-1} &=& \alpha_3^{-1}(M_Z)
+ \frac{b_3}{4\pi}\text{ln}\left(\frac{M_Z}{M_{GUT}}\right)^2,
\\
\alpha_{GUT}^{-1} &=& \alpha_{EM}^{-1}(M_Z)\text{sin}^2\theta_W(M_Z)
+ \frac{b_2}{4\pi}\text{ln}\left(\frac{M_Z}{M_{GUT}}\right)^2,
\\
\alpha_{GUT}^{-1} &=& \frac{3}{5}\alpha_{EM}^{-1}(M_Z)\text{cos}^2\theta_W(M_Z)
+ \frac{b_1}{4\pi}\text{ln}\left(\frac{M_Z}{M_{GUT}}\right)^2,
\end{eqnarray}
where $b_a\,(a=1,\,2,\,3)$ are the one-loop beta functions determined by the particle contents running in the loop (see Appendix~(\ref{b}) for more details), and we have taken into account
\begin{eqnarray}
e=g_2\text{sin}\theta_W=g_Y\text{cos}\theta_W=\sqrt{\frac{3}{5}}g_1\text{cos}\theta_W=\sqrt{4\pi\alpha_{EM}}.
\end{eqnarray}
We allow the error on the coupling unification to be no more than $5\%$: $|\alpha_i^{-1}-\alpha _1^{-1}| / \alpha_1^{-1} \leq 5\%$ for $i=2$ and 3, and take the central values of the following measured quantities~\cite{PDG-2015}:
\begin{eqnarray}\label{measured}
\begin{split}
M_Z &= (91.1880\pm 0.0020) \text{ GeV} ~,
\\
\alpha_3(M_Z) &= 0.1193\pm 0.0016 ~,
\\
\alpha_{EM}^{-1}(M_Z) &= 127.916\pm 0.015 ~,
\\
\sin^2\theta_W &= 0.22333\pm 0.00011 ~.
\end{split}
\end{eqnarray}
Under the unification conditions (\ref{gauge-coupling-unification}), (\ref{condi:perturbative}) and (\ref{condi:proton-decay}), and considering the fact that $b_i$ should not be smaller than the SM values: $b_1\geq \frac{41}{10},\,b_2\geq -\frac{19}{6},\,b_3\geq -7$, we can find constraints on $b_i$'s.

In the following, we constrain the new matter fields in Table~\ref{spectrum} and Table~\ref{spectrum-2} necessary for the gauge coupling unification, and assume that they have different numbers of generations $n_a$, where $a=Q,\,U,\,D,\,L,\,E,\,U'$ respectively.

\subsection{Supersymmetric case}

The supersymmetric spectrum of our models consists of the MSSM particles in the visible sector, new scalars and their vector-like superpartners charged under both the visible and the hidden groups, and the hidden vector supermultiplets. In the traditional SUSY $SU(5)$ GUT models~\cite{Barbieri:2016cnt}, SM gauge coupling unification is attained at $M_{\text{GUT}}=(2 - 3) \times 10^{16}\,\text{GeV}$, with the beta functions
\begin{eqnarray}
(b_1,\,b_2,\,b_3)_{MSSM}=\left(\frac{33}{5},\,1,\,-3\right),
\end{eqnarray}
where the masses of all the superpartners of the SM particles have been set at $1$ TeV for simplicity. These masses can be varied a little bit, a few hundred GeV to a few TeV, and do not affect the result much.

First, let's consider adding complete GUT multiplets to the spectrum. Suppose we add $n_5$ generations of $\bar{\textbf{5}}$, $n_{10}$ generations of $\textbf{10}$ and $n_{10'}$ generations of $\textbf{10'}$, where $\bar{\textbf{5}}$ and $\textbf{10'}$ are the extra particles given by Eqs.~(\ref{rep_5bar}) and (\ref{rep_10}), respectively, and $\textbf{10}$ is the 10-plet that incorporates the multiplet $\widetilde{U}^{\dagger}$ charged in Table~\ref{spectrum-2}.  We assume that the SUSY breaking scale is $5$ TeV.  Above $5$ TeV, the beta functions~\footnote{For definiteness, we take all the new scalars and the extra Higgs bosons to be at $300$ GeV, and all the new fermions and the superpartners of the SM to be at $5$ TeV. }
\begin{eqnarray}
\begin{split}
b_1 &= \frac{33}{5}+N(n_{5}+3n_{10}+3n_{10'}) ~, \\
b_2 &= 1+N(n_{5}+3n_{10}+3n_{10'}) ~, \\
b_3 &= -3+N(n_{5}+3n_{10}+3n_{10'}) ~,
\end{split}
\end{eqnarray}
where the factor of $N$  in the second terms comes from the fact that all the particles in Table~\ref{spectrum} are fundamental under the hidden group. The lower indices $10$ and $10'$ refer to the 10-plets involving $\widetilde{U}^{\dagger}$ and $\widetilde{U'}^{\dagger}$, respectively. As discussed earlier, we need the existence of multiplets $\Phi_{\widetilde Q}$ and $\Phi_{\widetilde U^{\dagger}}$ in order to have a Yukawa coupling to the Higgs in the superpotential, implying both $n_{10},\,n_{10'}\geq 1$. We find that adding at least $1$ generation of these two 10-plets to the otherwise (approximated) unified MSSM spectrum ruins the gauge coupling unification greatly due to the Landau pole. Even if we add incomplete GUT representations instead, the minimal requirement of at least one $\Phi_{\widetilde Q}$ and one $\Phi_{\widetilde U^{\dagger}}$ still blows up the running couplings at high energies. This reflects the fact that we have added too many particles to the low-energy (compared to the GUT scale) spectrum. This also indicates that our models favor a split-SUSY scenario, in which some of the SM superpartners and the new fermions are at the GUT scale. Note that here how exactly the spectrum splits is model-dependent, and is different from that in the standard split-SUSY scenarios where Higgsinos and gauginos are the lightest super partners.

For simplicity and in order not to add too many new particles due to their hidden multiplicities $N$, in the following we restrict our models to the $N=2$ case, though analysis shows that $N=3$ also works for achieving the SM gauge coupling unification.

\subsection{Non-supersymmetric case}

In this subsection, we consider the unification conditions when  only the new scalar parts of the multiplets charged under $SU(2)_H$ exist at low energies. This may be viewed as a decoupling limit of the non-traditional split-SUSY scenario (what we mentioned at the end of last subsection) where all the superpartners are at the GUT scale. We focus on the scenario where the new scalar masses are below the confinement scale $\Lambda_H$ and only new scalars in Table~\ref{spectrum} are taken into account. From now on we drop the prime notation in $\widetilde{U'}^{\dagger}$, since for the $SU(2)_H$ case the multiplet $\widetilde{U'}^{\dagger}$ identifies with the one without prime, $\widetilde{U}^{\dagger}$. For definiteness, we assume in this subsection that $\Lambda_H=4\,\text{TeV}$, that the masses of the extra Higgs bosons and all the new scalars are about $300$ GeV, and that those new scalars form bound states with masses around $800$ GeV or higher. We find the beta functions above the confinement scale $\Lambda_H$ to be
\begin{eqnarray}\label{bi}
\begin{split}
b_1 &= \frac{41}{10} +\frac{1}{20}+\frac{1}{5}\left(\frac{n_Q}{6}+\frac{4n_U}{3}+\frac{n_D}{3}+\frac{n_L}{2}+n_E\right) ~, \\
b_2 &= -\frac{19}{6}+\frac{1}{6}+\frac{n_Q}{2}+\frac{n_L}{6} ~, \\
b_3 &= -7+\frac{n_Q}{3}+\frac{n_U+n_D}{6} ~,
\end{split}
\end{eqnarray}
where the first term in each $b_a$ is purely the SM contributions, and the second terms in $b_1$ and $b_2$ come from the contributions of the additional Higgs degrees of freedom.

\begin{figure}[!htp]
\centering
\includegraphics[width=10cm]{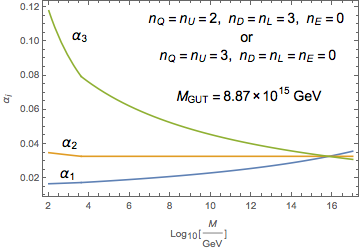}
\caption{Running gauge couplings of a non-supersymmetric scenario.  The new scalars and the extra Higgs bosons are taken to have mass at $300$ GeV.  Below the confinement scale $\Lambda_H=4\,\text{TeV}$, all the new scalars form bound states. }\label{fig:unification}
\end{figure}

We focus on one type of solutions that achieve the gauge coupling unification:
\begin{eqnarray}
\begin{split}
&
n_Q=n_U=2 ~,\,n_D=n_L=3 ~,\,n_E=0 ~,\quad \text{or} 
\\
&
n_Q=n_U=3 ~,\,n_D=n_L=n_E=0 ~,
\end{split}
 \label{non-susy-example}
\end{eqnarray}
with the unification scale $M_{GUT}\sim8.87\times 10^{15}$ GeV, as shown in Figure~\ref{fig:unification}.  One can easily check that the 1-loop beta function for $SU(2)_H$ is
\begin{eqnarray}
b_{H2}=-\frac{11}{3}\times 2+\frac{1}{6}\times 2\times \frac{1}{2}\times (3\times 3\times 2+3\times 3)=-\frac{22}{3}+4.5<0,
\end{eqnarray}
if there is no purely hidden matter field. Therefore, as long as there are not too many purely hidden degrees of freedom, the $SU(2)_H$ may be asymptotic free in the UV.

Below $\Lambda_H$, all the new scalars presumably form mesons (as listed in Table~\ref{mesons}).  Therefore, they do not contribute to the beta functions.  One can check that even if a baryon with nontrivial $SU(2)_L$ and $U(1)_Y$ charges (see Table~\ref{baryons}) is formed below $\Lambda_H$, its contributions to running couplings are negligible, due to the small value of $\Delta b_i\times \text{ln}\,\frac{4\,\text{TeV}}{800\,\text{GeV}}$~\footnote{Note that above the QCD confinement scale and below the hidden confinement scale, the new particles are within the SM perturbative but the hidden non-perturbative regime. The hidden binding force contributes to the masses of the bound states. For definiteness, we assume that all new bound states have masses at $800$ GeV.}. The threshold corrections induced by non-perturbative effects around the hidden confinement scale are neglected.

In the example of Eq.~(\ref{non-susy-example}), the new scalars in the spectrum form incomplete GUT representations.
The existence of incomplete GUT representations at low energies is typical in four-dimensional (4D) GUT theories. One famous example is the Higgs doublet-triplet splitting problem, namely, in the splitting
\begin{eqnarray}
\label{5}
\textbf{5} \to \left( \textbf{3},\,\textbf{1},\,-\frac{1}{3} \right)
+ \left( \textbf{1},\,\textbf{2},\,\frac{1}{2} \right) ~,\\
\label{5bar}
\bar{\textbf{5}}\to \left( \bar{\textbf{3}},\,\textbf{1},\,\frac{1}{3} \right)
+ \left( \textbf{1},\,\textbf{2},\,-\frac{1}{2} \right) ~.
\end{eqnarray}
The colored triplets are heavy while the $SU(2)_L$ doublets are light. This splitting may be related to the $\mu$ problem. The doublets and the triplets attain masses via a superpotential with coupling to an adjoint field and a $\mu$ term upon the GUT symmetry breaking:
\begin{eqnarray}
W_5&=&\lambda \,\bar{\textbf{5}}\cdot \textbf{24}\cdot \textbf{5}+\mu \,\bar{\textbf{5}}\cdot\textbf{5}\notag\\
&\Rightarrow&\left(2\lambda v+\mu\right)\bar{\textbf{3}}\cdot\textbf{3}
+ \left(-3\lambda v+\mu\right)\bar{\textbf{2}}\cdot\textbf{2} ~.
\end{eqnarray}
With $v$ around the GUT scale and doublets at $\mathcal O(100)$ GeV, a tuning for the $\mu$ parameter is needed. In the case of $SO(10)$ GUT, one way to explain the doublet-triplet splitting is via the Dimopoulos-Wilczek mechanism~\cite{GUT-pheno:DRW82}. Generally speaking, the strategy to generate such mass splitting in a GUT multiplet is to construct a superpotential in such a way that some components of the multiplet get masses around the GUT scale and thus decouple from the low-energy spectrum. However, it requires a careful arrangement of the VEV's of the other fields (in particular singlets) in the superpotential and is usually complicated. 

The mass splitting issue may also be explored in the context of extra spacial dimensions with orbifold~\footnote{For a pedagogical review, see for instance Ref.~\cite{SUSY-GUT:R08}.}. The idea is that fields localized at the fixed points of the internal space survive the orbifold actions and remain as complete multiplets in 4D, while fields living in the bulk are partially projected out and thus form incomplete multiplets~\footnote{Such a scenario commonly appears in local GUT models in which the SM gauge symmetry arises as intersections of several larger symmetries at different orbifold fixed points~\cite{local-GUT:BLS07}.}. Whether a GUT multiplet lives in the bulk or at a fixed point is model-dependent. The masses that the bulk fields acquire are not arbitrary, even in the case of extra scalars and vector-like fermions. They must be induced by the VEV's of some auxiliary singlets~\footnote{In string models, these massive bulk fields must satisfy the string selection rules~\cite{string-selection-rules}. }.

We assume that there exists an underlying higher dimensional theory that generates the mass splitting in the example of Eq.~(\ref{non-susy-example}) and leave the explicit construction of an orbifold model giving such a spectrum to a future work.

\section{Phenomenological Searches and Bounds }\label{sec:bounds-bound-states}

\subsection{Bounds from Colliders and Precision Observables}\label{sec:bounds}

As alluded to earlier, we have assumed that the hidden strong dynamics with an $\mathcal O (1)$-TeV confinement scale has a much shorter hadronization time scale than QCD to ensure the new scalars to form bound states before they decay. The conventional collider constraints on the R-hadrons may not completely apply to our models, since in addition to QCD the hidden force plays a more dominant role in bound state formation. Direct searches of the hidden scalars are also very different from those of squarks/sleptons in the MSSM. However, we still have some indirect bounds coming from the electroweak precision constraints of LEP experiments since most of our new particles have the SM charges and may couple to the SM Higgs boson. The bounds on the electroweak $S$, $T$, $W$, $Y$ parameters due to the hidden scalars are similar to the supersymmetry precision bounds given in Ref.~\cite{Marandella:2005wc}. New hidden scalars heavier than a couple of hundred GeV are safe from the constraints.  These bounds can be further relaxed by decoupling the hidden scalars from the SM Higgs boson, which does not change the phenomenology of our bound states much.

Another indirect bound comes from the Higgs data, since the hidden scalars running in the loops will modify Higgs production and branching ratios, mostly constrained from the $H \rightarrow \gamma\gamma$ and $H\rightarrow gg$ channels.  Again, hidden scalars heavier than $300$ GeV are safe~\cite{Cheung:2015uia}.  Although in our models the hidden scalars come from the hidden chiral multiplets as extensions to the MSSM, those indirect bounds still apply.
\begin{figure}
\begin{center}
\includegraphics[width=8cm]{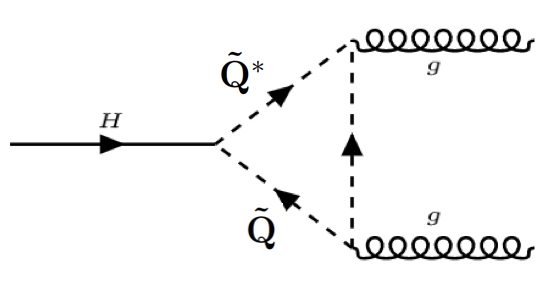}
\includegraphics[width=8cm]{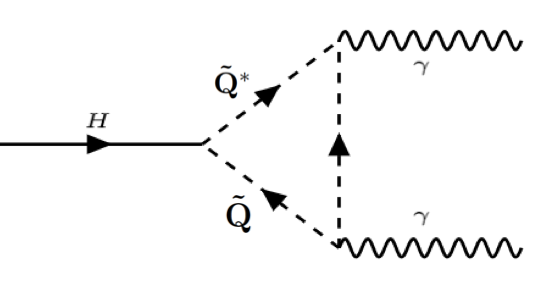}
\caption{The scalar quarks from hidden sector can modify the Higgs production rate and branching ratios through the loops. }\label{Hprecision}
\end{center}
\end{figure}

\subsection{Exotic Bound States}\label{sec:bound-states}

The hidden scalars (and possibly with the SM particles) form exotic bound states under $SU(2)_H$.  For now, we only focus on the lightest bound states with 2 matter particles, and assume that all the vector-like fermions, if they exist, have mass of at least a few TeV~\footnote{The vector-like fermions in our models can form bound states just like their scalar superpartners, and the mixed bound states formed by both fermions and scalars can be present too.  However, under the assumption that the hidden fermions are heavier, all the bound states involving vector-like fermions are at least around a few TeV and thus beyond our current consideration.}. 

To find out all the possible bound states, we list in Table~\ref{reps} relevant products of irreducible representations under different gauge groups that contain singlets under SM and hidden strong interactions.
\begin{table}[h]
\centering
\begin{tabular}{cc}
\hline \hline
 $SU(2)$ & $\textbf{2}\otimes \textbf{2}$ \\
 \hline
  $SU(3)$ & $\textbf{3}\otimes \bar{\textbf{3}}$, $\textbf{3}\otimes \textbf{3}\otimes \textbf{3}$ \\
 \hline\hline
 \end{tabular}
\caption{Products of irreducible representations that contain singlets. The conjugate representations are not listed.}\label{reps}
\end{table}

\subsubsection{Exotic Mesons}

The exotic mesons are of the type $A A^{\dagger}$, listed in Table~\ref{mesons}, where $A$ refers to an exotic particle from Table~\ref{spectrum} and the dagger indicates its conjugate. They are neutral under both visible and hidden gauge groups. 
\begin{table}[h]
\centering
\begin{tabular}{c}
\hline\hline
Exotic Mesons 
\\
\hline
\vspace{-0.5cm}
\\
 $\widetilde{Q}\widetilde{Q}^{\dagger}$, $\widetilde{U}\widetilde{U}^{\dagger}$, $\widetilde{D}\widetilde{D}^{\dagger}$, $\widetilde{L}\widetilde{L}^{\dagger}$, $\widetilde{E}\widetilde{E}^{\dagger}$
 \\
 \hline\hline
\end{tabular}
\caption{Exotic mesons formed from hidden scalars, all neutral under both SM and hidden gauge groups.}\label{mesons}
\end{table}

We assume that the masses of the lightest hidden scalars do not exceed the hidden confinement scale $\Lambda_H$.
Compared to the SM mesons, among which the lightest one is CP-odd, the lightest composite state in our models is expected to be a CP-even neutral meson instead, as a result of an S-wave bound state of the messenger scalars.

The supersymmetric setup can be compared to models with fermionic bi-fundamental constituents ({\it e.g.}, composite/little Higgs models), where the lightest singlet appears as a Goldstone boson mode.  In those models, one of the neutral Goldstone bosons becomes heavier than the SM-charged Goldstone bosons due to the chiral anomaly of the hidden gauge interaction. The neutral composite states in our scenario can be lighter due to mixing among the exotic mesons and possible the hidden glueball, which can couple to the SM gauge bosons through scalar-loop diagrams (see also Ref.~\cite{Novikov:1977dq}).

\subsubsection{Exotic Baryons}

Under $SU(2)_H$, the exotic baryons that consist of 2 matter particles have the form $AA'$, where $A$ and $A'$ denote distinct hidden scalars (or their conjugates) ({\it i.e.}, $A' \neq A^{\dagger}$). They are listed in Tables~\ref{baryons}.  
\begin{table}[h]
\centering
\begin{tabular}{cccc}
\hline\hline
$AA'$ & $SU(2)_L$&$U(1)_Y$ & $U(1)_{EM}$\\
\hline
$\widetilde{Q}\widetilde{U}^{\dagger}$ & $\textbf{2}$ & $-\frac{1}{2}$ & $-1,\,0$
\\
$\widetilde{Q}\widetilde{D}^{\dagger}$ & $\textbf{2}$ &$\frac{1}{2}$ &$0,\,1$
\\
$\widetilde{L}\widetilde{E}^{\dagger}$ & $\textbf{2}$&$\frac{1}{2}$ & $0,\,1$
\\
$\widetilde{L}\widetilde{E}$ & $\textbf{2}$ &$-\frac{3}{2}$& $-2,\,-1$
\\
$\widetilde{U}\widetilde{D}^{\dagger}$ &$\textbf{1}$ & $1$ &$1$
\\
\hline\hline
\end{tabular}
\caption{Exotic baryons as $SU(2)_L$ doublets and singlets and their Abelian charges. The conjugate particles are not listed.}\label{baryons}
\end{table}

We point out that different than the QCD and composite-Higgs models, the baryon masses in our models are not correlated with the confinement scale $\Lambda_H$. This is because the baryons in our models are constructed by complex scalars instead of (approximate) chiral fermions.

We briefly mention the existence of the $AA'a$ type of exotic baryon states, where $A$ and $A'$ are new hidden scalars (or their conjugates) and $a$ refers to a SM quark.  One example is the bound state $\widetilde{Q}\widetilde{Q}u_R$, which is a singlet or triplet under $SU(2)_L$, taking hypercharge $1$ and electric charge $0$, $\pm 1$, or $\pm 2$. For these bound states, $AA'$ forms an $SU(2)_H$ singlet with a nontrivial QCD charge, and then forms a bound state with a SM quark by the QCD strong force. Since the QCD confinement scale $\Lambda_{QCD}\sim \mathcal{O} (100)\,\text{MeV}$ is much lower than the hidden confinement scale $\Lambda_H\sim\mathcal O (1)\,\text{TeV}$, we expect the $AA'a$ bound states to be much more unstable than the $AA'$ baryons. Since the $AA'a$-type baryons have various $SU(2)_L$ charges (singlet, doublet, triplet, or quadraplet), their decay channels can be quite interesting~\footnote{Their decays may require the existence of additional particles to mediate the interactions.}. Bounds for long-lived R-hadrons formed from colored SUSY particles in exotic or split SUSY models may apply here~\cite{Aad:2011yf}. Bound states of $\mathcal O(1)$ TeV in mass are generally safe from such constraints. We leave the analysis of their interactions and decay signatures at colliders to a future work.

\subsection{Dark Matter Candidates}

In this subsection, we explore the existence of a dark matter (DM) candidate among the exotic baryon states formed by the messenger fields. For the $SU(2)_H$ gauge group, the $AA'$-type baryons may or may not be stable, depending on whether there is a symmetry (or topology such that the exotic baryons can be interpreted as solitons) to ensure their stability. If such a symmetry exists, {\it e.g.}, a hidden baryon number $U(1)_H$ that gives each particle in Table~\ref{spectrum} a hidden baryon number $1/2$, then the first three and the last baryons in Table~\ref{baryons} have hidden baryon number $1$ while the fourth one has hidden baryon number 0.
Then the lightest, electrically neutral of the first three baryons can be a DM candidate, and the discussion of the corresponding relic abundance follows the line of Refs.~\cite{hidden-strong-dynamics:CFHIY1507,hidden-strong-dynamics:CIY1512}. In that case, the DM baryons annihilate into a pair of lighter scalar non-baryonic composite states.  The thermal relic abundance can be much lower than the observed DM density if the annihilation cross section (into mesons, glubeballs, etc) saturates the unitarity bound~\cite{DM-unitarity}. 
As a rough estimate, the relic abundance of an $AA'$-type DM baryon is~\cite{hidden-strong-dynamics:CIY1512}
\begin{eqnarray}\label{DM_abund}
\Omega _Bh^2\sim  10^{-5}\frac{1}{F(M_B)^4}\left(\frac{M_B}{1\,\text{TeV}}\right)^2,
\end{eqnarray}
where $M_B$ is the mass of the DM baryon, denoted as $B$, and $F(M_B)$ is the form factor of the interaction of the DM baryon with lighter states such that $F=1$ when the unitarity limit is saturated. 
With the lightest new hidden scalar mass being $\sim 300$~GeV, we expect that the lightest exotic baryon and the lightest exotic meson should have roughly the same mass in the sub-TeV regime.  As an example that will be discussed in the next section, we have $\widetilde U$ and $\widetilde Q$ significantly lighter than the other hidden scalars, with the former slightly lighter than the latter.  In this case, the DM baryon would be slightly heavier than the lightest exotic meson.
We note from (\ref{DM_abund}) that as long as the form factor $F(M_B)$ is not smaller than a certain value (i.e. $F(M_B)\gtrsim \left[\frac{10^{-5}}{0.12}\left(\frac{M_B}{1\,\text{TeV}}\right)^2\right]^{1/4}\approx 0.096\left(\frac{M_B}{1\,\text{TeV}}\right)^{1/2}$ for cold, non-baryonic DM), the DM baryon would not overclose the universe. A more exact calculation of the relic abundance requires a detailed knowledge of the hidden strong dynamics, particularly the precise form of the form factor. 

The direct search of the DM candidate will be through the couplings between the hidden scalars and the Higgs boson~\footnote{The $Z$ exchange for electrically neutral bound states may also contribute and face roughly the same constraints as estimated in Ref.~\cite{GUTzilla-DM:HLL2016}.}.
In terms of the effective field theory (EFT) approach, that corresponds to the direct coupling between the DM baryon and the Higgs boson
\begin{eqnarray}
\mathcal L\ni \lambda _B B^{\dagger}BH^{\dagger}H,
\end{eqnarray}
where the coupling is of the same order as the couplings between the constituent new scalars and the Higgs boson. The corresponding DM elastic interaction with nuclei via the Higgs exchange will lead to a spin-independent (SI) cross section~\cite{DM:KMNO2010} 
\begin{eqnarray}
\sigma _{SI}=\frac{\lambda_B^2}{4\pi m_h^4}\frac{m_N^4f_N^2}{M_B^2}\approx 1.36\times 10^{-44}\text{cm}^2\times \lambda_B^2\left(\frac{1\,\text{TeV}}{M_B}\right)^2,
\end{eqnarray}
where we have used the lattice result of the nucleon decay constant $f_N\approx 0.326$~\cite{lattice-fN-value09} in the SM, and $m_N$ is the nucleon mass at around 1 GeV.  An exact estimate depends on the DM mass and the coupling $\lambda_B$ that encodes effects of the hidden strong dynamics. It seems that this cross section can satisfy the current LUX limit~\cite{LUX2015} $\sigma_{SI}\lesssim 1\times 10^{-44}\text{cm}^2(M_B/1\,\text{TeV})$ and be within the reach of the proposed LUX-Zeplin (LZ) experiment~\cite{LZ2015} for a suitable value of $\lambda_B$.  However, for an baryon with nonvanishing hypercharge, its interaction with nucleons through Z-boson can be dominant. In fact, the cross section of such a process reads~\cite{YZBLYZ2011}
\begin{eqnarray}
\sigma _{SI}^{Z} = \frac{m_N^2M_B^2}{\pi (M_B+m_N)^2}\left(\frac{F_N}{\sqrt{2}}\right)^2,
\end{eqnarray}
where $F_N$ is an induced form factor with mass dimension $-2$:
\begin{eqnarray}
F_p=2F_u+F_d\quad\text{for proton},\quad\quad F_n=F_u+2F_d\quad\text{for neutron},
\end{eqnarray}
with form factors $F_{u/d}\sim\mathcal{O}(1)/M_Z^2$. One finds that $\sigma_{SI}^Z\sim\mathcal{O}(10^{-35})\,\text{cm}^2$, which is much above the LUX limit. This shows that a DM candidate does not exist in the lowest exotic baryons formed by the messenger fields listed in Table~\ref{spectrum} for the $SU(2)_H$ gauge group. When one considers an $SU(3)_H$ hidden gauge sector instead of $SU(2)_H$, such dangerous Z-boson exchange interactions between a baryon and a nucleon can be turned off due to the existence of baryon states (consisting of $3$ messenger scalars) neutral (which we call ``completely neutral") under both $SU(2)_L$ and $U(1)_Y$ (e.g. $\widetilde{L}_N\widetilde{L}_E\widetilde{E}^{\dagger}$). The SI scatterings with nuclei for such baryons are dominated by the Higgs exchange and thus their cross sections can satisfy the LUX limit and be within the reach of the LZ experiment. The lightest one of these completely neutral baryons can be a DM candidate.

If there is no symmetry or topology to ensure the stability of the exotic baryons, they will decay~\footnote{In this case, probably additional particles need to be present for the decays to occur.}. We leave the  study of the decay patterns of the exotic baryons in the follow-up work.

\section{Collider phenomenology}\label{sec:750}

The various bound states have rich phenomenology in colliders, with different masses and decay channels. Thus the most significant signatures of our models are a tower of resonances with different masses. 
Although any model with hidden strong dynamics or in the context of GUT's can predict such resonances as well, we emphasize that our models are quite different from the composite-Higgs/technicolor scenarios due to the existence of fundamental scalars and different types of fundamental degrees of freedom, as mentioned in Section~\ref{sec:intro}.

As the simplest and lightest bound states, here we focus on the phenomenology of  lightest  exotic mesons at the LHC. 
We generically denote a scalar meson by $S$ with mass $M_S$ and a pseudoscalar meson by $P$ with mass $M_P$.  One distinctive property between these two types of exotic mesons, for example, is that the latter cannot decay into a pair of Higgs bosons while the former can as long as kinematics allows.

As argued in the previous section, the lightest meson in our models has to be CP-even.  We use an EFT approach to describe the dynamics of the bound states at low energies, similar to the formalism of pion interactions in strongly-coupled QCD.  The detailed calculation is summarized in Appendix~\ref{app:EFT}.  
For definiteness and simplification, we make the following assumptions for the hidden scalar masses:
\begin{itemize}
\item Both $\widetilde Q$ and $\widetilde U$ representations are respectively degenerate, {\it i.e.},
\begin{eqnarray}
m_{\widetilde Q_1}=m_{\widetilde Q_2}=... ~,\quad m_{\widetilde U_1}=m_{\widetilde U_2}=... ~,
\end{eqnarray}
where the lower indices of $\widetilde Q$ and $\widetilde U$ denote generations.
\item Both $\widetilde Q$ and $\widetilde U$ have similar masses and are lighter than the other scalars, {\it i.e.},
\begin{eqnarray}
m_{\widetilde{Q}}\approx m_{\widetilde{U}}<\text{ mass of other hidden scalars} 
\label{non-dengeneracy}.
\end{eqnarray}
\end{itemize}

\begin{figure}[!htp]
\includegraphics[width=8cm]{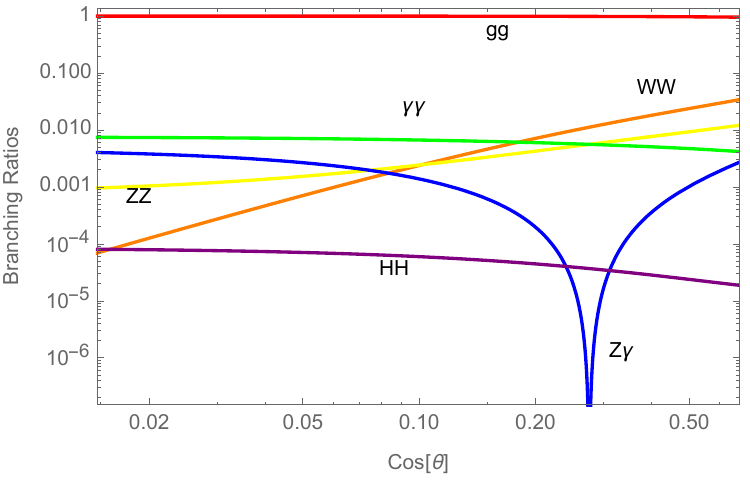}
\includegraphics[width=8cm]{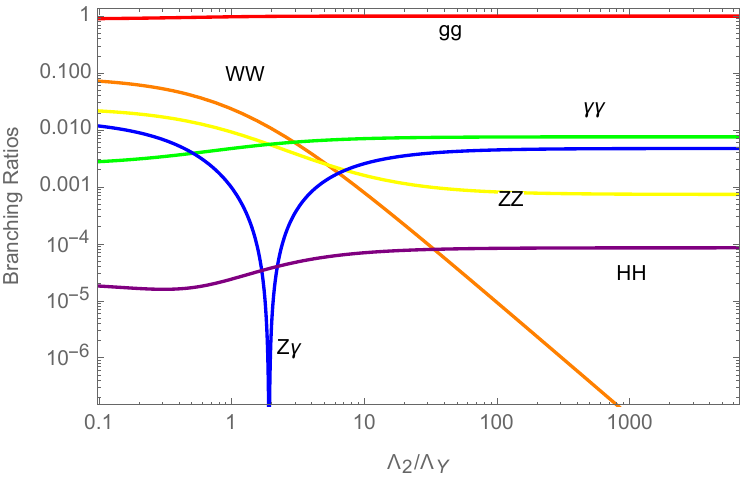}
\caption{Branching ratios of the lightest scalar resonance of mass $M_S = 2$~TeV decaying into the gauge boson pairs and the Higgs boson pair as functions of $\text{cos}\,\theta$ (left) and $\Lambda_2/\Lambda_Y$ (right) for $\Lambda_{H} = 4$~TeV. Both left and right panels are depicted for $\lambda=0.01$.}
\label{fig:branching}
\end{figure}

Under these assumptions, it is justifiable to consider the glue-glue fusion (GGF) as the dominant production process for the exotic mesons~\footnote{For the 13-TeV proton-proton collisions, gluons have a larger parton distribution function (PDF) than those of quarks, and thus contribute dominantly in the production of the resonance, as long as the couplings of the resonance to $W,\,Z$ bosons and the photon are not too large.  Figure~\ref{fig:branching} justifies that in the specific example considered here, the GGF process indeed dominates over the vector-boson fusion and the photon fusion processes.}.  Note that Eq.~(\ref{non-dengeneracy}) follows one of the examples with satisfactory gauge coupling unification given in Eq.~(\ref{non-susy-example}).

As an explicit example, we consider a scalar resonance $S$ of mass 2~TeV.
Figure~\ref{fig:branching} shows the dependence of branching ratios of $S$ on $\cos\theta$ (left plot) and $\Lambda_2/\Lambda_Y$ (right plot).  The mixing angle $\theta$  is defined in Eq.~(\ref{mixing-angle}) to denote the rotation angle when one converts the meson flavor eigenstates to the mass eigenstates.  The suppression scales $\Lambda_i$ ($i=Y,\,2,\,3$) encoding details of hidden strong dynamics are defined in Eq.~(\ref{suppressionscales}).  From Eq.~(\ref{suppression-scales}), one finds the relation
\begin{eqnarray}
8|\tan\theta|=\biggl|\frac{15\Lambda_2}{\Lambda_Y}-1\biggr|.
\end{eqnarray}
In the numerical analysis here, we use the running gauge coupling constants evaluated at the renormalization scale of $M_S$, where  new particles involved in the coupling running are taken as in Eq.~(\ref{non-susy-example}).  The cusps at $\text{cos}\,\theta\approx 0.27$ and $\Lambda_2/\Lambda_Y\approx 2$ for the $Z\gamma$ curves reflect the fact that the $Z\gamma$ decay width vanishes if $\frac{g_2^2}{\Lambda_2}=\frac{5}{3}\frac{g_Y^2}{\Lambda_Y}$,  due to a destructive interference effect as seen in Eq.~(\ref{decay-width:Zgamma}).  As $S$ is purely composed of colored particles, the decay rate of $S$ into gluons is preponderant, particularly when its effective coupling to the SM Higgs boson, $\lambda$ defined in Eq.~(\ref{lambda}), is diminishing.  In these plots, we take $\lambda = 0.01$.  Therefore, the exotic mesons are dominantly produced via the GGF process.  The branching ratio of the $HH$ channel, proportional to $\lambda^2$, is thus subdominant in most of the parameter space.  For simplicity and definiteness, we will neglect the effects of the operator in Eq.~(\ref{eq:Higgs}) by assuming $\lambda \to 0$, so that the di-Higgs bound~\cite{bound:diHiggs} is trivially satisfied.

\begin{figure}[!htp]
\includegraphics[width=8cm]{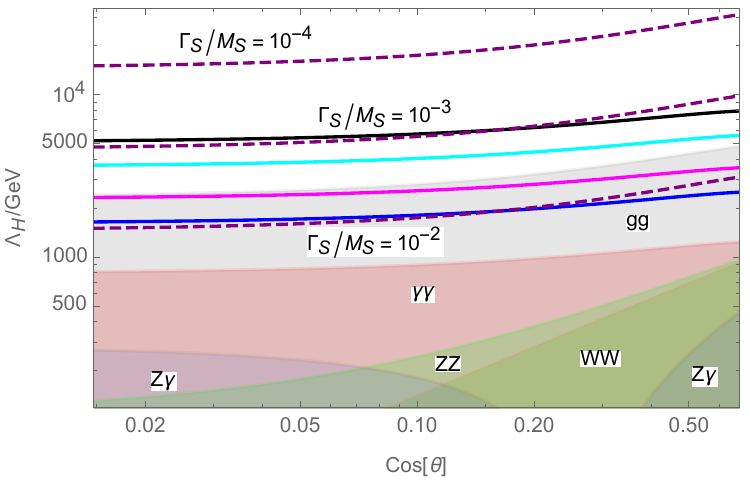}
\includegraphics[width=8cm]{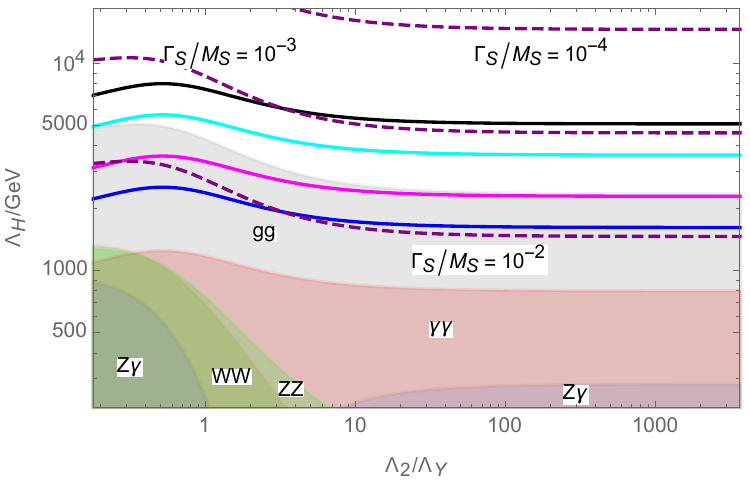}
\caption{Contours of the production cross section of $S$ times its decay branching ratio into $\g\g$ at the $13$-TeV LHC by assuming the glue-glue fusion production process.  We fix $M_S = 2$~TeV and take the factorization and renormalization scales at $\mu = M_S/2$.  The dashed purple curves are contours for specific values of $\Gamma_S/M_S$.  The shaded regions are excluded by the searches through various decay modes in LHC Run-I and Run-II as detailed in Eq.~(\ref{g}).  The black, cyan, magenta and blue solid lines correspond to $ \sigma _{\gamma\gamma}^{(13\,{\rm TeV})} = 0.1,\,0.2,\,0.5,\,1$~fb, respectively.}
\label{fig:crosssection}
\end{figure}

Even though with a small branching ratio, as shown in Figure~\ref{fig:branching}, it is expected that such a new scalar will probably first show up through the diphoton channel due to the clean signals at the LHC.
Figure~\ref{fig:crosssection} shows the contours of the diphoton channel cross section at the 13-TeV LHC, $\sigma_{\gamma\gamma}^{(13\,{\rm TeV})}$, on the $\cos\theta$-$\Lambda_H$ plane (left plot) and the $\Lambda_2/\Lambda_Y$-$\Lambda_H$ plane for a few representative values of $\Gamma_S/M_S$.  Note that here we use $M_S/2$ as the factorization and renormalization scale when calculating the cross sections.
The colored regions are excluded by the resonance searches using various decay channels in LHC Run-I and Run-II
for $M_S=2$~TeV:
\begin{align}
\begin{split}
\sigma(pp\to S \to \g\g) &< 0.3\,{\rm fb}\quad  \mbox{\cite{CMS:2014onr, Aad:2015mna}} \ ,\\
\sigma(pp\to S \to WW) &< 60\,{\rm fb}\quad  \mbox{\cite{Khachatryan:2015cwa, Aad:2015agg,ATLAS-CONF-2016-074}}\ , \\
\sigma(pp\to S \to ZZ) &< 8\,{\rm fb}\, \quad \mbox{\cite{Aad:2015kna,CMS-PAS-B2G-16-023}}\ , \\
\sigma(pp\to S \to Z\gamma) &< 20\, {\rm fb}\quad \mbox{\cite{Aad:2014fha,CMS-PAS-EXO-17-005}}\ , \\
\label{g}\sigma(pp\to S \to jj) &< 70\,{\rm pb}\quad  \mbox{\cite{Aad:2014aqa,CMS:2015neg,CERN-EP-2017-042,Khachatryan:2015sja}} \ . 
\end{split}
\end{align}
The figure shows that certain parameter space predicts $\s_{\g\g}^{(13\,{\rm TeV})} = {\cal O}(0.1)$~fb
while evading all the above constraints for an $\mathcal O(1)$~TeV hidden confinement scale
\begin{eqnarray}
\L_{H}  \simeq 4~{\rm TeV} ~,
\end{eqnarray}
with the relative suppression scale, $\Lambda_2/\Lambda_Y \agt 1$ or $\text{cos}\,\theta \alt 0.5$. 
This result implies that $\widetilde{U}^{\dagger}$ is slightly lighter than  $\widetilde{Q}$ in the specific example. Figure~\ref{fig:crosssection} (which reveals information about $\Lambda_H$, messenger mass hierarchy, etc) reflects the fact that the messenger resonances serve as a good probe to the new physics. Moreover, $\Gamma_S/M_S \sim {\cal O}(10^{-3})$, justifying our narrow width approximation employed in numerical analyses. 

We now comment on the cross sections of the resonance decaying into other channels.  Different than the scenario in Ref.~\cite{hidden-strong-dynamics:CIY1512}, for the specific example considered here the branching ratios of the $WW$, $ZZ$, $Z\gamma$ modes change dramatically, as shown in Figure~\ref{fig:branching}.  Instead of a simple, approximate proportionality relation between the cross sections of the other gauge boson modes and that of the diphoton mode, the ratios $\sigma (pp\to S\to WW/ZZ/Z\gamma/jj)\,/\,\sigma(pp\to S\to \gamma\gamma)$ depend on $\cos\theta$ or $\Lambda_2/\Lambda_Y$ in a more complicated way.  In addition, compared to Figure~2 of Ref.~\cite{hidden-strong-dynamics:CIY1512}, here the contours of the di-photon cross sections in Figure~\ref{fig:crosssection} vary less drastically as $\text{cos}\,\theta$ or the relative suppression scale $\Lambda_2/\Lambda_Y$ changes. This is because in Ref.~\cite{hidden-strong-dynamics:CIY1512} the mixing is between one lepton-like and one down-type quark-like scalars, while in our specific example the resonance is made of only quark-like scalars.  Such differences reflect the fact that the particle contents in the resonance meson greatly affect how the meson decays.  With future data on the other decay channels available, we will know better about the nature of particles involved in the lowest-lying resonance mesons.  This may also shed light on the mass hierarchy in the hidden sector. Through such searches, the resonance structure may be revealed.

\section{Summary}\label{sec:concl}

We explore a class of models consisting of the Standard Model (SM) and a strongly coupled hidden $SU(N)_H$ sector.  These models have various interesting phenomenological properties.  First, by adding at low energies new messenger (hidden) fields taking special charges under the SM and the hidden gauge groups, the SM gauge couplings may unify at a suitable scale without inducing the fast proton decay issue.  Their charge assignments are chosen in such a way to be embedded into a larger gauge group $SU(5)_V\times SU(N)_H$. Analysis shows that a small-rank $SU(N)_H$ ($N=2,\,3$) gauge group and a split-SUSY scenario are favored. 
Secondly, the existence of new scalars provides an alternative solution to the electroweak hierarchy problem with or without SUSY.  The Higgs mass can thus be less fine-tuned due to the cancellation between the top loop and the hidden scalar loop contributions.
Thirdly, different from little-Higgs/composite-Higgs/technicolor models, the Higgs fields in our models are fundamental particles whose electroweak symmetry breaking pattern largely remains intact at low energy scales, indicating that our models suffer less from collider and electroweak precision observable constraints.  Further studies of properties of the new exotic resonances, such as branching ratios to the SM particles, will distinguish our models from composite/little Higgs models.  Finally, the models predict many exotic bound states formed under the hidden strong dynamics and the SM strong force. The bound states can cover a wide range of spectrum, some of which are within the reach of the LHC in the near future. They may also provide us with DM candidates with various weak charges and spins, which have not been widely discussed in the literature yet.  We leave the comprehensive study of the stability of new bound states and their decay signatures to a future work.

The potentially unified gauge theory and the strong hidden dynamics probably show some hints on the existence of an underling theory. In fact our models may be well implemented into string theory. The strongly coupled hidden $SU(N)_H$ sector can arise from the world volume gauge field living on a stack of D-branes, and become part of the near throat strong warping AdS background geometry in the AdS/CFT limit, while the visible perturbative $SU(5)_V$ sector can originate from another stack of D-branes that intersect with the hidden branes (and do not warp the throat). Open string modes at the intersections attaching to the two stacks of branes as bi-fundamentals under both $SU(5)_V$ and $SU(N)_H$ gauge groups play the role of messengers between the visible and the hidden sectors. In this sense, our models may be viewed as a field theory realization of the so-called holographic gauge mediation~\cite{holographic-gauge-mediation:MSS}, where the hidden $SU(N)_H$ is studied in the large $N$ limit as in holography.

Back to 1974~\cite{Kamenik:2016izk}, it was an exciting time for the community to first realize QCD as the gauge theory for strong interactions and to study the bound states of quarks.  The LHC opens a new era in particle physics.  With QCD already established, it is tantalizing to speculate there may be additional strongly coupled sectors in Nature.  Such new strong dynamics may arise in different ways.  They can be classified by whether they participate in the electroweak symmetry breaking and whether the new fundamental degrees of freedom are fermionic or bosonic.  Through careful searches of their decay signatures in high-energy collisions, we may soon reveal the hidden strong dynamics at the LHC.

\acknowledgments

S.S. and F.Y. would like to thank the hospitality of NCTS where this work was initiated. S.S. would like to thank Henry Tye, Ann Nelson and Haipeng An for useful discussions about supersymmetry. F.Y. would like to thank Gary Shiu and Eibun Senaha for helpful discussions about gauge coupling unification.  This work was supported in part by the Ministry of Science and Technology of Taiwan under Grant Nos. MOST104-2628-M-002-014-MY4 and MOST104-2811-M-008-043 and in part by the CRF Grants of the Government of the Hong Kong SAR under HUKST4/CRF/13G.

\appendix

\section{One-Loop Beta Functions}\label{app:beta}

At one-loop level, the beta functions of SM gauge coupling constants read
\begin{eqnarray}\label{running-coupling}
\alpha_a^{-1}(\mu)=\alpha^{-1}_a(\mu_0)+\frac{b_a}{4\pi}\text{ln}\left(\frac{\mu_0^2}{\mu^2}\right),
\end{eqnarray}
where $a=1,\,2,\,3$ refer to gauge groups $U(1)_Y$, $SU(2)_L$ and $SU(3)_C$, respectively. 
The group-theoretic coefficients $b_a$ depend on the numbers of particles running in the loop.
\begin{eqnarray}\label{b}
b_a=-\frac{11}{3}\sum_VC(R_V^a)+\frac{2}{3}\sum_{Weyl}C(R_F^a)+\frac{1}{6}\sum_{Real}C(R_S^a) ~,
\end{eqnarray} 
where $R_V^a$, $R_F^a$ and $R_S^a$ refer to vector, Weyl fermion and real scalar representations under the gauge group labeled by $a$, respectively. $C(R^a)$'s are Dynkin indices defined through
\begin{eqnarray}
\text{Tr}\left(T^A_RT^B_R\right)=C(R)\delta^{AB}
\end{eqnarray}
for the non-Abelian group representation $R$. We choose the normalization such that for the fundamental representation $\textbf{N}$ under $SU(N)$, the Dynkin index is
\begin{eqnarray}
C(\textbf{N})=\frac{1}{2} ~.
\end{eqnarray}
Then under $SU(N)$, the Dynkin indices for the adjoint representation, the asymmetric tensor with rank 2, and the symmetric tensor with rank 2 are
\begin{eqnarray}
C(a)=N,\quad C(A_2)=\frac{N-2}{2},\quad C(S_2)=\frac{N+2}{2},
\end{eqnarray}
respectively.  Eq.~(\ref{b}) can be applied to the Abelian group $U(1)_Y$ as well, by replacing $C(R_V^a)\to 0$, $C(R_F^a)\to \frac{3}{5} Y_F^2$ and $C(R_S^a)\to \frac{3}{5}Y_S^2$, where $Y_i$'s are hypercharges.
The renormalization factor $\frac{3}{5}$ is introduced as the hypercharge is identified with the diagonal generator of $SU(5)$ that does not belong to the Cartan subalgebras of $SU(3)$ and $SU(2)$
\begin{eqnarray}
T^{24}=\sqrt{\frac{3}{5}} Y ~.
\end{eqnarray}

\section{Effective Field Theory for the Bound States}\label{app:EFT}

Effective interactions of a scalar $S$ or a pseudoscalar $P$ with the SM gauge bosons can be parametrized by 
\begin{eqnarray}
\label{eq:SEFT}
{\cal L}_{\rm eff} ^S=
 \frac{\kappa_3 ^{(S)}}{\Lambda_H}S G^a_{\mu\nu}G^{a\,\mu\nu}
 + 
  \frac{\kappa_2^{(S)}}{\Lambda_H}S W^i_{\mu\nu}W^{i\,\mu\nu}
 + 
\frac{5}{3} \frac{\kappa_Y^{(S)} }{\Lambda_H}S B_{\mu\nu}B^{\mu\nu}\ ,
\end{eqnarray}
\begin{eqnarray}
\label{eq:PEFT}
{\cal L}_{\rm eff}^P=
 \frac{\kappa_3 ^{(P)}}{\Lambda_H}P \widetilde{G}^a_{\mu\nu}G^{a\,\mu\nu}
 + 
  \frac{\kappa_2^{(P)}}{\Lambda_H}P \widetilde{W}^i_{\mu\nu}W^{i\,\mu\nu}
 + 
\frac{5}{3} \frac{\kappa_Y^{(P)}}{\Lambda_H}P \widetilde{B}_{\mu\nu}B^{\mu\nu}\ ,
\end{eqnarray}
where $\Lambda_H$ is the emergent hidden strong dynamical scale.  Here, $G^{\mu\nu}$, $W^{\mu\nu}$ and $B^{\mu\nu}$ denote the field strengths of the SM gauge bosons of the $SU(3)_C$, $SU(2)_L$, and $U(1)_Y$ groups, respectively, with the superscripts $a$ and $i$ being the indices for the corresponding adjoint representations. The coefficients $\kappa_{3,2,1}^{(S/P)}$ are $S/P$-dependent ${\cal O}(1)$ parameters and encapsulate 
details of the strong dynamics. One may define suppression scales for a canonically normalized $S/P$ coupling to different gauge groups:
\begin{eqnarray}\label{suppressionscales}
\frac{1}{\Lambda_i}=\frac{\kappa_i^{(S/P)}}{\Lambda_H},\quad i=Y,\,2,\,3.
\end{eqnarray}

The kinetic terms of the gauge fields are 
\begin{eqnarray}
{\cal L}=- \frac{1}{4g_3^2}  G^a_{\mu\nu}G^{a\,\mu\nu} 
- \frac{1}{4g_2^2} W^i_{\mu\nu}W^{i\,\mu\nu} 
-\frac{1}{4g_Y^{2}} B_{\mu\nu}B^{\mu\nu}\ ,
\end{eqnarray}
where $g_3$, $g_3$ and $g_Y$ are the corresponding gauge couplings. 

We assume any new complex scalar to be massive~\footnote{We have assumed that the corresponding hidden fermions, if they exist, are at lease a few TeV and much heavier than the hidden scalars and thus decouple from the low-energy effective theory.}:
\begin{eqnarray}
\label{eq:mass}
{\cal L} \supset -m_{Qi}^2 \widetilde{Q}_i^\dagger \widetilde{Q}_i ~,
\end{eqnarray}
where $\widetilde{Q}_i$ here refers to any new particles in Table~\ref{spectrum}, the mass $m_{Qi}\gtrsim 300$ GeV, consistent with the current bounds from electroweak precision observables and Higgs measurements as discussed in Section~\ref{sec:bounds}.  It can be even lighter if some of the new bi-fundamental scalars have suppressed couplings to the SM Higgs boson.

The lightest meson made of the new scalars is kinetically normalized as 
\begin{eqnarray}\label{S}
S=\frac{4\pi}{\kappa \Lambda_h}\sum _i O_{1i}[\widetilde{Q}^{\dagger}_i\widetilde{Q}_i] ~,
\end{eqnarray}
where the bracket refers to a meson state, $4\pi$ is introduced through Naive Dimensional Analysis (NDA)~\cite{Cohen:1997rt,Luty:1997fk}. The matrix $O=(O_{ij})$ is a special orthogonal matrix that brings the mesons to their mass eigenstates, and the scalar $S$ corresponds to the first (lowest) eigenstate. The mesons in the original basis is
\begin{eqnarray}\label{mesons-original}
[\widetilde{Q}_i^{\dagger}\widetilde{Q}_i]=\frac{\kappa\Lambda_H}{4\pi}\,O_{1i}\,S ~.
\end{eqnarray}
Plugging (\ref{mesons-original}) back into the effective lagrangian with the mesons in the original basis, we can read off the suppression scales $\Lambda_1$, $\Lambda_2$ and $\Lambda_3$ as functions of $O_{1i}$'s respectively.

In order to highlight the relation between the branching ratios of the lightest meson $S$ and the mixings of particles forming $S$, let us assume the flavors within the $\widetilde Q$ and $\widetilde U$  multiplets are degenerate and
\begin{eqnarray}
m_{\widetilde{Q}}\approx m_{\widetilde{U}}<\text{ mass of any other new scalars} \label{App:non-dengeneracy}.
\end{eqnarray}
Note that Eq.~(\ref{App:non-dengeneracy}) may be viewed as a spacial case of the example shown in Eq.~(\ref{non-susy-example}). 


Under the above assumptions, the meson mass eigenstates $S$ and $T$ (with $M_S<M_T$) can be obtained through an $SO(2)$ rotation parametrized by an angle $\theta$ and a rescaling
\begin{eqnarray}\label{mixing-angle}
\left(\begin{array}{c}
S\\
T\end{array}\right)
&=&\frac{4\pi}{\kappa \Lambda_H}\left(\begin{array}{cc}
\text{cos}\,\theta & \text{sin}\,\theta\\
-\text{sin}\,\theta & \text{cos}\,\theta\end{array}\right)
\left(\begin{array}{c}
\widetilde{Q}^{\dagger}\widetilde{Q}\\
\widetilde{U}^{\dagger}\widetilde{U}\end{array}\right),
\end{eqnarray}
where $\kappa \sim \mathcal{O}(1)$. The effective Lagrangian of the mass eigenstates is
\begin{eqnarray}
\mathcal L_{eff}&\ni&\frac{\kappa}{4\pi\Lambda_H}\biggl[\bigl(2\text{cos}\,\theta+\text{sin}\,\theta\bigr)SG^2+3\text{cos}\,\theta\,SW^2
+\bigl(\frac{1}{3}\,\text{cos}\,\theta+\frac{8}{3}\,\text{sin}\,\theta\bigr)SB^2\biggr]+... ~,
\end{eqnarray}
where the ellipses refer to the couplings between the heavier meson $T$ and gauge bosons. Thus we can read the suppression scales
\begin{eqnarray}\label{suppression-scales}
\begin{split}
\frac{1}{\Lambda_3} &= \frac{\kappa\bigl(2\text{cos}\,\theta+\text{sin}\,\theta\bigr)}{4\pi\Lambda_H} ~, \\
\frac{1}{\Lambda_2} &= \frac{3\kappa\,\text{cos}\,\theta}{4\pi\Lambda_H} ~,\\
\frac{1}{\Lambda_Y} &= \frac{3}{5}\frac{\kappa}{4\pi\Lambda_H} 
\left(\frac{1}{3}\,\text{cos}\,\theta+\frac{8}{3}\,\text{sin}\,\theta \right) ~.
\end{split}
\end{eqnarray}

Through the effective couplings with the gluons and in the narrow width approximation, the scalar/pseudoscalar resonance is produced at the LHC via the gluon fusion process 
\begin{eqnarray}
\begin{split}
\sigma(pp \to S/P) &=
\frac{\pi^2}{8} \left(\frac{\Gamma(S/P\to g+g)}{M_{S/P}}\right) \times 
\frac{1}{s}\frac{\q{\cal L}_{gg}}{\q\cal \t}
~,
\\
\frac{\q{\cal L}_{gg}}{\q\cal \t} &= \int_0dx_1 dx_2 f_g(x_1) f_g(x_2) \delta (x_1 x_2 - \tau)\ ,
\end{split}
\end{eqnarray}
where $\tau = M_{S/P}^2/s$ and $\sqrt{s}$ denote the center-of-mass energy of the proton-proton collisions.
Using the MSTW PDF's~\cite{Martin:2009iq}, we obtain  
\begin{align}
\frac{1}{s}\frac{\q{\cal L}_{gg}}{\q\cal \t} &\simeq
\left\{\begin{array}{ll}
1.1~{\rm pb} & ({\rm for } \sqrt{s} = 8~{\rm TeV})\ ,\\
15~{\rm pb} & ({\rm for } \sqrt{s} = 13~{\rm TeV})\ ,
\end{array}\right.
\end{align}
where we have fixed the factorization scale and the renormalization scale at $\mu = M_S/2$ for $M_S = 2$~TeV.

The partial decay widths of the scalar resonance are given by
\begin{eqnarray}
\label{eq:SGG}
\G(S/P\to gg) &=& \frac{2}{\pi} \left( \frac{g_3^2}{\Lambda_3}  \right)^2 M_{S/P}^3 ~,
\\
\label{eq:SWW}
\G(S/P\to W^+ W^-) &=&  
 \frac{1}{2\pi} \left( \frac{g_2^2}{\Lambda_2} \right)^2 M_{S/P}^3 ~,
 \\
\label{eq:SZZ}
 \G(S/P\to ZZ) &=& 
 \frac{1}{4\pi} \left[
 \left( \frac{g_2^2}{\Lambda_2}\right)  c_W^2
+
\frac{5}{3}\left(
\frac{g_Y^{ 2}}{\Lambda_Y} 
\right)s_W^2 
 \right]^2 M_{S/P}^3 ~, 
\\
 \G(S/P\to \g\g) &=& \frac{1}{4} 
 \frac{1}{\pi} \left[ 
  \left( \frac{g_2^2}{\Lambda_2}\right)  s_W^2
+
\frac{5}{3}\left(
\frac{g_Y^{ 2}}{\Lambda_Y} 
\right)
c_W^2
 \right]^2 M_{S/P}^3 ~, 
 \\
\label{decay-width:Zgamma}
\G(S/P\to Z\g) &=& 
 \frac{1}{2\pi} \left[ 
  \left( \frac{g_2^2}{\Lambda_2}\right)  
-
\frac{5}{3}\left(
\frac{g_Y^{ 2}}{\Lambda_Y} 
\right)
 \right]^2
 c_W^2s_W^2
 M_{S/P}^3 ~, 
\end{eqnarray}
where $s_W \equiv \sin\theta_W$ and $c_W = (1-s_W^2)^{1/2}$ with $\theta_W$ being the weak mixing angle. 
The masses of the $W$ and $Z$ bosons are neglected to a good approximation.



The decay of $S$ into a pair of the $125$-GeV Higgs bosons is characterized by interactions between new particles and the Higgs boson $H$,
\begin{eqnarray}\label{lambda_i}
{\cal L} = \left(\lambda _Q\, \widetilde{Q}^\dagger \widetilde{Q}\,+\lambda_U\,\widetilde{U}^{\dagger}\widetilde{U}\right) H^\dagger H \ ,
\end{eqnarray}
with $\l_{Q,U}$ being coupling constants.
These interactions induce an effective interaction between $S$ and Higgs doublets,
\begin{eqnarray}
\label{eq:Higgs}\label{lambda}
 {\cal L} =  \frac{\l}{4\pi} \L_{H}S H^\dagger H\ , 
\end{eqnarray}
where we again use the NDA and reparameterize $\l_{Q,U}$ and $\theta$ by $\lambda$.
Through this operator, the resonance  decays into a pair of Higgs bosons with a partial decay width~\footnote{Strictly speaking, the operator in Eq.~(\ref{eq:Higgs}) also induces the decays into the weak gauge bosons at loop levels.}:
\begin{eqnarray}
\G(S\to HH^\dagger) = \frac{1}{8\pi M_S}\left( \frac{\lambda \L_H}{4\pi}\right)^2 \ .
\end{eqnarray}
Notice that the pseudoscalar resonance will not decay to di-Higgs bosons.



\end{document}